\documentclass[iop,floatfix,numberedappendix,twocolappendix]{emulateapj}

\usepackage[backref,breaklinks,colorlinks,urlcolor=blue,citecolor=blue,linkcolor=blue]{hyperref}
\usepackage[all]{hypcap}

\usepackage{graphicx}
\usepackage{enumerate}
\usepackage{bm}
\usepackage{color}
\usepackage[utf8]{inputenc}

\begin{document}

\title{A Disk-based Dynamical Constraint on the Mass of the Young Binary DQ Tau}
\author{I.~Czekala\altaffilmark{1}, S.~M.~Andrews\altaffilmark{1}, G.~Torres\altaffilmark{1}, E.~L.~N.~Jensen\altaffilmark{2}, K.~G.~Stassun\altaffilmark{3,4}, D.~J.~Wilner\altaffilmark{1} \& D.~W.~Latham\altaffilmark{1}}
\altaffiltext{1}{Harvard-Smithsonian Center for Astrophysics, 
			 60 Garden Street, Cambridge, MA 02138; \email{iczekala@cfa.harvard.edu}}
\altaffiltext{2}{Department of Physics and Astronomy, Swarthmore College, 500 College Avenue, Swarthmore, PA 19081}
\altaffiltext{3}{Department of Physics and Astronomy, Vanderbilt University, Nashville, TN 37235}
\altaffiltext{4}{Department of Physics, Fisk University, Nashville, TN 37208}

\newcommand{\teff}{T_{\rm eff}}
\newcommand{\logg}{\log g}
\newcommand{\radmc}{\texttt{RADMC-3D}}
\newcommand{\kms}{\textrm{km~s}^{-1}}
\newcommand{\todo}[1]{\textcolor{red}{#1}}
\newcommand{\vt}{ {\bm \theta}}
\newcommand{\msun}{M_\odot}

\begin{abstract}
We present new ALMA observations of CO $J$=2$-$1 line emission from the DQ~Tau circumbinary disk. These data are used to tomographically reconstruct the Keplerian disk velocity field in a forward-modeling inference framework, and thereby provide a dynamical constraint on the mass of the DQ~Tau binary of $M_\ast = 1.27_{-0.27}^{+0.46} \,M_\odot$.  Those results are compared with an updated and improved orbital solution for this double-lined system based on long-term monitoring of its stellar radial velocities.  Both of these independent dynamical constraints on the binary mass are in excellent agreement: taken together, they demonstrate that the DQ~Tau system mass is $1.21\pm0.26\,M_\odot$ and that the disk and binary orbital planes are aligned within 3\degr\ (at 3\,$\sigma$ confidence). The predictions of various theoretical models for pre-main sequence stellar evolution are also consistent with these masses, although more detailed comparisons are difficult due to lingering uncertainties in the photospheric properties of the individual components.  DQ~Tau is the third nearly equal-mass double-lined spectroscopic binary with a circumbinary disk that has been dynamically ``weighed" with these two independent techniques: all show consistent results, validating the overall accuracy of the disk-based approach and demonstrating that it can be robustly applied to large samples of young, {\it single} stars as ALMA ramps up to operations at full capacity.     
\end{abstract}

\section{Introduction \label{sec:intro}}

Theoretical models for pre-main sequence (pre-MS) stellar evolution are fundamental tools for learning about star and planet formation.  But the accuracy of such models -- especially at young ages -- is unclear, due to our limited understanding of some complex physical effects like accretion \citep[e.g.,][]{baraffe09} or magnetic fields and convection \citep[e.g.,][]{feiden13}.  While such issues are being explored theoretically, robust observational constraints on key stellar parameters can be used to help guide improvements to the models.  Most important are dynamical constraints on stellar masses, $M_{\ast}$ \citep[e.g., see][]{hillenbrand04}.  Usually these are determined from the orbital motions of binary systems \citep{stassun14}, but they could increasingly be measured for {\it single} stars based on the rotation of their associated gas disks \citep[e.g.,][]{simon00}.      

We are interested in comparing the constraints from these two approaches, to illuminate and quantify any associated systematic problems in the inference of $M_{\ast}$.  To do that, we have targeted the few roughly equal-mass double-lined spectroscopic binaries that host circum-binary disks, including V4046 Sgr \citep{rosenfeld12} and AK Sco \citep{czekala15}.  In both cases, excellent agreement (to within $\sim$1\%) is found between the estimates of $M_{\ast}$ from radial velocity monitoring of the stars and the tomographic reconstruction of the CO gas velocity field in the disk.  The confluence of these measurements also indicates that the disk and binary orbital planes are well-aligned (within 1--2\degr).  Moreover, the predictions of theoretical pre-MS models faithfully reproduce these results for these two particular examples.  However, these model successes are perhaps not surprising, since both V4046 Sgr and AK Sco are relatively old (10 and 18\,Myr) and massive (1.8 and 2.5\,$M_{\odot}$ in each system) and the models should be more robust in that range of age and mass.  An important supplementary test would employ a cooler and younger binary.      

In those respects, DQ~Tau is an exemplary target.  DQ~Tau is a roughly equal-mass double-lined spectroscopic binary with a period of $\sim$16 days and a notably eccentric orbit \citep{mathieu97}.  It has a composite spectral type of $\sim$M0--M1 \citep[e.g.,][]{herbig77,herczeg14} and is located in the nearby and relatively young Taurus clouds. DQ Tau exhibits enhancements of various tracers of accretion and activity --- optical brightening \citep{mathieu97}, emission line variations \citep{basri97}, and mm/radio emission \citep{salter10} --- that have been associated with both pulsed accretion and reconnection events from colliding magnetospheres near peri-astron.  The DQ~Tau binary hosts a circumbinary disk with substantial millimeter continuum emission from dust \citep{beckwith90,andrews05,guilloteau11}.  There is recent evidence for molecular gas in rotation around the central binary host \citep{williams14}, although there is non-negligible contamination from the local molecular cloud \citep{guilloteau13}.         

We present new observations of molecular gas in the DQ~Tau circumbinary disk made with the Atacama Large Millimeter/Submillimeter Array (ALMA), and use them to place a dynamical constraint on the total mass of the DQ~Tau binary.  We also provide an updated orbital solution for the DQ~Tau binary based on long-term radial velocity monitoring.  Section~\ref{sec:data} presents the data and its calibration.  Section~\ref{sec:method} describes our modeling of the gas disk velocity field, provides an update of the original \citet{mathieu97} orbital solution, and highlights the key results.  Section~\ref{sec:disc} discusses these results together and assesses the predictions of pre-MS evolution models.  And Section~\ref{sec:summary} provides a summary in the contexts of other young circumbinary disk systems and the utility of the disk-based dynamical mass technique.

\section{Observations and Data Reduction}\label{sec:data}

\subsection{Millimeter Interferometry}

ALMA observed the DQ Tau system on 2015 May 24, using 34 of its 12\,m antennas with separations ranging from 21 to 540\,m.  The observations were configured with the same spectral setup as in \citet{czekala15}, employing the Band 6 receivers to cover the CO, $^{13}$CO, and C$^{18}$O $J$=2$-$1 transitions in 61, 61, and 122\,kHz channels, respectively, as well as the adjacent continuum (at 232\,GHz, or 1.3\,mm).  The nearby quasar J0510+1800 (6\degr\ separation) was observed regularly to monitor variations in the complex gain response of the interferometer.  The bright quasar J0423$-$0120 was also observed to calibrate the bandpass behavior and absolute flux levels.  The total on-source integration time was $\sim$28 minutes.  The visibilities were calibrated using standard techniques with the {\tt CASA} software package (v4.3).  After a phase and amplitude self-calibration based on the bright continuum, the spectral line visibilities were time-averaged (to 30\,s intervals) and continuum-subtracted.        

Images of the continuum and spectral line data were created by Fourier inverting the calibrated visibilities (assuming a Briggs robust weighting parameter of 0.5, to balance S/N and resolution), deconvolving with the {\tt CLEAN} algorithm, and restoring with a synthesized beam with FWHM = $0\farcs8\times0\farcs6$ (at P.A. = 145\degr).  The continuum image shows a bright, marginally resolved source with a peak intensity of $\sim$68\,mJy beam$^{-1}$ and integrated flux density of 79\,mJy.  The RMS noise level is 70\,$\mu$Jy beam$^{-1}$ (the peak S/N is $\sim$1000; the map sensitivity is clearly limited by dynamic range).  All of this emission is expected to be from dust; the peri-apse continuum enhancement noted by \citet{salter10} is not present at the observed orbital phase ($\phi = 0.71$).  But the focus here is on the emission from the CO spectral lines.  

\begin{figure*}[htb]
\begin{center}
  \includegraphics{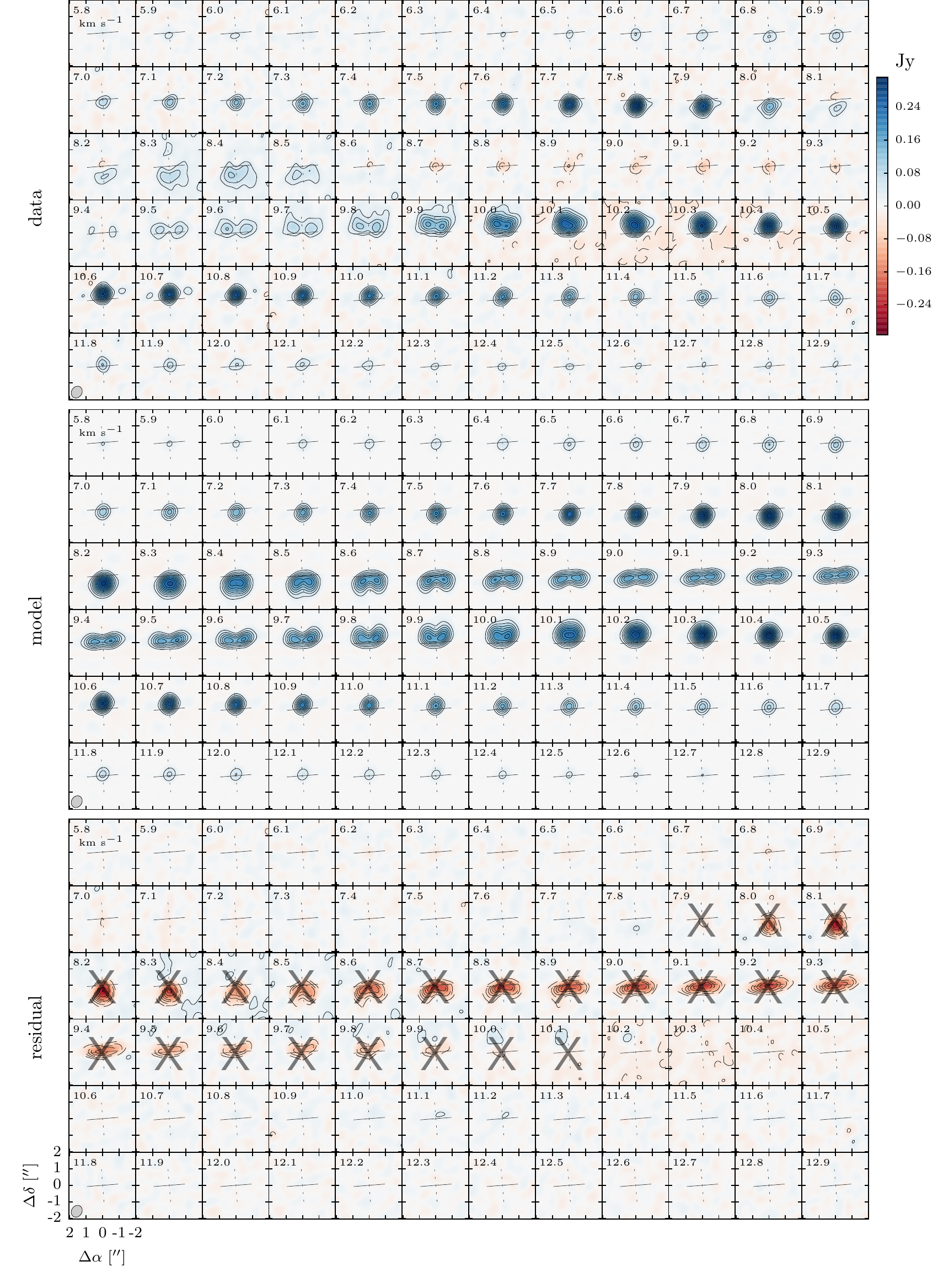}
  \figcaption{
  CO $J$=2$-$1 channel maps for the DQ~Tau data ({\it top}), the best-fit model ({\it middle}), and the imaged residuals ({\it bottom}) at 0.1\,$\kms$ velocity resolution.  Contours are drawn at intervals of 3$\times$ the RMS noise level (9.5\,mJy beam$^{-1}$).  The synthesized beam is drawn in the lower left corner of each set of channel maps, and the LSRK velocities are labeled in each panel.  We do not model the channels with $v = 7.9 - 10.1\,\kms$ (inclusive) because of cloud contamination; these channel maps are marked with an $\mathsf{X}$ in the lower panel.  
  \label{fig:results}}
  \end{center}
\end{figure*}

The $^{12}$CO (hereafter CO) and $^{13}$CO $J$=2$-$1 transitions were imaged in 0.1 km s$^{-1}$-wide channels, and reach an RMS noise level of 8\,mJy beam$^{-1}$ in each channel.  Line emission from these transitions is detected over a $\sim$7 km s$^{-1}$-wide velocity range, exhibiting the classical morphological pattern of Keplerian rotation.  The peak S/N is 35 for CO, but only 7 for $^{13}$CO.  Both of these transitions show considerable contamination from the local molecular cloud material, affecting a 2 km s$^{-1}$-wide span slightly blueshifted from the systemic velocity.  The C$^{18}$O $J$=2$-$1 transition was imaged in 0.2 km s$^{-1}$-wide channels, with an RMS of $\sim$4\,mJy beam$^{-1}$ in each, but the line is only marginally detected (S/N$\sim$3) in a few of these channels.  Given the line intensities, our focus will be on an interpretation of the CO emission.  The CO channel maps are shown in the top portion of Figure~\ref{fig:results}.

\subsection{Optical Spectroscopy}

Three sets of optical spectroscopic observations, including material also used by \cite{mathieu97}, were used to re-examine the orbital solution of the DQ~Tau binary. The first set consists of 30 spectra obtained at the Harvard-Smithsonian Center for Astrophysics (CfA) between 1984 and 2005 with two similar instruments equipped with intensified photon-counting Reticon detectors, as described in more detail by the above authors. These spectrographs are no longer in operation. A subset of 23 of these spectra was used by \cite{mathieu97}; we have reanalyzed all 30 of them here with improved techniques. These single-order spectra (45\,\AA\ centered around the \ion{Mg}{1}\,b triplet near 5190\,\AA) have relatively low S/N, ranging from 6 to 16 per 8.5~$\kms$ resolution element. A second set of 22 observations consists of radial velocity differences measured from spectra also described by \cite{mathieu97} and collected with instruments at the Lick Observatory, the Keck Observatory, and the McDonald Observatory.  Finally, more recently (2013 October to December) we obtained three additional spectra of DQ~Tau at the CfA for a different purpose, with the 1.5\,m Tillinghast reflector at the Fred L.\ Whipple Observatory on Mount Hopkins (AZ). For this we used the bench-mounted TRES instrument \citep{furesz08} that delivers a resolving power of $R \approx 44,000$ in 51 echelle orders spanning the wavelength range 3900--9100\,\AA. These three spectra have signal-to-noise ratios in the \ion{Mg}{1}\,b region of 16, 26, and 23 per 6.8~$\kms$ resolution element.

\capstartfalse
\begin{deluxetable*}{lc rcc rcc}
\tablewidth{0pc}
\tablecaption{Heliocentric radial velocity measurements of DQ~Tau from CfA.\label{table:RVs}}
\tablehead{
\colhead{HJD} &
\colhead{Orbital} &
\colhead{$RV_1$} &
\colhead{$\sigma_1$} &
\colhead{$(O-C)_1$} &
\colhead{$RV_2$} &
\colhead{$\sigma_2$} &
\colhead{$(O-C)_2$}
\\
\colhead{(2,400,000$+$)} &
\colhead{phase\tablenotemark{a}} &
\colhead{($\kms$)} &
\colhead{($\kms$)} &
\colhead{($\kms$)} &
\colhead{($\kms$)} &
\colhead{($\kms$)} &
\colhead{($\kms$)}
}
\startdata
    45982.0293 &   0.1435 &  34.76   &  3.93 &  $-$2.71  &  12.44  &  2.53 &  $+$1.75  \\
    46389.8446 &   0.9520 & $-$2.93  &  3.61 &  $-$0.57  &  56.13  &  2.32 &  $+$2.89  \\
    46745.8122 &   0.4794 &  37.11   &  7.72 &  $+$6.50  &  13.90  &  4.97 &  $-$4.12  \\
    47073.9364 &   0.2446 &  37.38   &  3.56 &  $+$0.49  &  15.01  &  2.29 &  $+$3.69  \\
    47075.9910 &   0.3747 &  34.73   &  3.35 &  $+$0.97  &  17.06  &  2.16 &  $+$2.41  \\
    47078.0084 &   0.5023 &  31.64   &  3.23 &  $+$1.79  &  20.70  &  2.08 &  $+$1.87  \\
    47127.8232 &   0.6549 &  26.07   &  4.56 &  $+$2.22  &  24.40  &  2.94 &  $-$0.84  \\
    47128.8081 &   0.7172 &  23.51   &  5.11 &  $+$2.88  &  27.30  &  3.29 &  $-$1.38  \\
    47159.6132 &   0.6667 &  20.10   &  4.96 &  $-$3.19  &  25.73  &  3.19 &  $-$0.11  \\
    47159.6318 &   0.6679 &  17.31   &  4.46 &  $-$5.92  &  27.86  &  2.87 &  $+$1.96  \\
    47427.9265 &   0.6468 &  27.28   &  3.19 &  $+$3.06  &  23.67  &  2.06 &  $-$1.17  \\
    47546.6122 &   0.1579 &  36.99   &  5.66 &  $-$0.66  &   7.64  &  3.64 &  $-$2.86  \\
    47546.6272 &   0.1588 &  33.17   &  5.28 &  $-$4.48  &  10.94  &  3.40 &  $+$0.44  \\
    47546.6272 &   0.1588 &  35.11   &  3.15 &  $-$2.54  &   9.87  &  2.02 &  $-$0.63  \\
    47546.6423 &   0.1598 &  35.24   &  6.15 &  $-$2.42  &  10.05  &  3.96 &  $-$0.44  \\
    47789.9412 &   0.5569 &  28.81   &  2.92 &  $+$0.89  &  19.93  &  1.88 &  $-$0.96  \\
    47840.8048 &   0.7758 &  12.43   &  3.01 &  $-$4.45  &  33.26  &  1.94 &  $+$0.57  \\
    47845.8709 &   0.0964 &  40.09   &  2.92 &  $+$4.81  &  13.83  &  1.88 &  $+$0.79  \\
    47896.7312 &   0.3151 &  35.14   &  3.12 &  $-$0.19  &  11.47  &  2.00 &  $-$1.51  \\
    47898.7686 &   0.4440 &  35.89   &  3.31 &  $+$4.16  &  16.98  &  2.13 &  $+$0.16  \\
    47902.6961 &   0.6926 &  19.12   &  3.35 &  $-$2.86  &  26.86  &  2.16 &  $-$0.38  \\
    48525.9580 &   0.1356 &  37.08   &  3.27 &  $-$0.22  &  10.20  &  2.10 &  $-$0.67  \\
    48670.6445 &   0.2920 &  36.83   &  3.41 &  $+$0.95  &  14.36  &  2.19 &  $+$1.97  \\
    48871.9572 &   0.0321 &  12.75   &  3.41 &  $-$7.74  &  23.86  &  2.19 &  $-$4.97  \\
    49344.8584 &   0.9596 & $-$6.27  &  4.17 &  $-$3.48  &  50.94  &  2.69 &  $-$2.76  \\
    49373.6908 &   0.7842 &  14.82   &  4.56 &  $-$1.44  &  37.61  &  2.94 &  $+$4.25  \\
    49410.6311 &   0.1220 &  34.84   &  4.56 &  $-$2.02  &   7.51  &  2.94 &  $-$3.83  \\
    49644.8270 &   0.9430 &   3.16   &  3.08 &  $+$4.65  &  52.75  &  1.98 &  $+$0.43  \\
    49699.8045 &   0.4223 &  32.37   &  3.08 &  $-$0.02  &  17.20  &  1.98 &  $+$1.08  \\
    53693.8806 &   0.1867 &  39.09   &  4.82 &  $+$1.44  &   8.27  &  3.10 &  $-$2.23  \\
    56578.9704 &   0.7691 &  17.46   &  3.95 &  $+$0.11  &  34.61  &  3.41 &  $+$2.43  \\
    56606.9106 &   0.5373 &  22.67   &  3.95 &  $-$5.97  &  26.25  &  3.41 &  $+$6.12  \\
    56650.8257 &   0.3164 &  31.34   &  3.95 &  $-$3.96  &  11.49  &  3.41 &  $-$1.52
\enddata
\tablenotetext{a}{Computed from the ephemeris given in Table~\ref{table:orbit}.}
\end{deluxetable*}
\capstarttrue

Radial velocities (RVs) for each component of DQ~Tau were measured from all of the CfA spectra using the two-dimensional cross-correlation technique TODCOR \citep{zucker94}, with templates taken from a library of synthetic spectra based on PHOENIX model atmospheres \citep[see][]{husser13} computed for the appropriate instrumental resolution. Based on indications from the work of \cite{mathieu97} that the mass ratio is close to unity, we assumed the stars have the same temperature.  Adopting solar metallicity, we experimented with templates of fixed surface gravities from $\log g = 3.5$ to 4.5.  The best matches to the DQ~Tau spectra were found for temperatures of $\sim$4000\,K (although a relatively wide range of $\pm$300\,K around that value is permissible) and $v \sin i$ values of 14 and 11~$\kms$\ for the primary and secondary, respectively. These latter values are similar to the measurements of \cite{nguyen12}, who obtained $14.7 \pm 1.6$ and $11.3 \pm 0.7$~$\kms$. 

The nominal temperature corresponds formally to a spectral type of K7, although $\teff$\ is degenerate with the surface gravity in our procedure because of the short wavelength coverage of the Reticon spectra from which we made these determinations. However, the RVs are unaffected by this degeneracy so long as the temperature for the templates is chosen to provide the optimal match to the spectra for a given $\log g$ value. Rotational broadening has a much larger effect on the velocities in this case because of the heavy line blending, and we believe our fine-tuning of this parameter for both stars is the reason we are able to derive meaningful velocities from all 23 of the spectra used by \cite{mathieu97} (in addition to the other 10 from CfA used here). Their procedures only allowed them to derive separate velocities for 14 of their least blended spectra, the rest providing only an upper limit of 18~$\kms$\ on the velocity separation between the primary and secondary.  We list our new radial-velocity measurements from all CfA spectra in Table~\ref{table:RVs}. 

The velocity zero-point of the Reticon observations was monitored each night by means of dusk and dawn exposures of the twilight sky, and small run-to-run corrections were applied in the manner described by \cite{latham92}. For TRES the zero point was monitored by observing IAU velocity standards each night. All velocities were placed on the same system, and the measurements listed in Table~\ref{table:RVs} include all corrections.

\section{Analysis and Results} \label{sec:method}

\subsection{CO Disk Modeling} \label{sec:disk}

We analyze the CO $J$=2$-$1 line emission using the framework detailed in \citet{czekala15} and \citet{rosenfeld12}.  Briefly, we forward-model the ALMA visibilities using a parametric description of the disk structure (densities, temperatures, and velocities).  For any set of model parameters, we calculate the excitation conditions in the disk assuming local thermodynamic equilibrium.  We then use the {\tt RADMC-3D} radiative transfer code \citep[v0.38;][]{dullemond12} to ray-trace spectral images, which are Fourier transformed and sampled at the same spatial frequencies as the data.  A $\chi^2$ likelihood function is used to assess the fit quality.  The posterior parameter-space is explored with a Markov Chain Monte Carlo (MCMC) algorithm.\footnote{The code used to perform the analysis described here is open source and freely available under an MIT license at \url{https://github.com/iancze/JudithExcalibur}} 

The model parameters can be catalogued into four groups.  The first group includes parameters that describe the CO gas densities.  We assume the standard \citet{lynden-bell74} similarity solution describes the radial surface density profile of the gas, which is described by an index $p$,\footnote{This is more commonly $\gamma$, but we aim to avoid confusion with the standard terminology in the RV analysis (see Sect.~\ref{sec:orbit}).} a characteristic radius $r_c$, and a normalization that we cast in terms of the CO gas mass $M_{\rm CO}$.  For computational expediency, we fix $p = 1$.  The second group describes the gas temperatures.  We simplify the scenario by assuming a vertically isothermal structure, with a radial power-law temperature profile that has normalization $T_{10}$ (the temperature at 10\,AU) and index $q$.  This thermal structure is employed in calculating the vertical density distribution, assuming hydrostatic equilibrium.  The third group of parameters sets the projected velocity field of the gas, presumed to be in Keplerian rotation.  It includes the central binary mass $M_\ast$, the disk inclination $i_d$ and position angle $\varphi$, and a systemic velocity $v_{\rm sys}$.  Non-thermal line broadening is permitted with a line-width $\xi$ added in quadrature to the normal thermal contribution.  In our convention, $i_d = 0\degr$ is a face-on disk with the angular momentum vector pointed towards the observer, $i_d = 90\degr$ is edge-on, and $i_d=180\degr$ is face-on but with the disk angular momentum vector pointed away from the observer.  The position angle $\varphi$ is defined by the projection of the angular momentum axis onto the sky.  The fourth group of parameters is utilitarian, including the distance $d$ and nuisance offsets from the observed phase center ($\Delta_{\alpha}$, $\Delta_{\delta}$).  

\begin{figure}[htb]
  \includegraphics{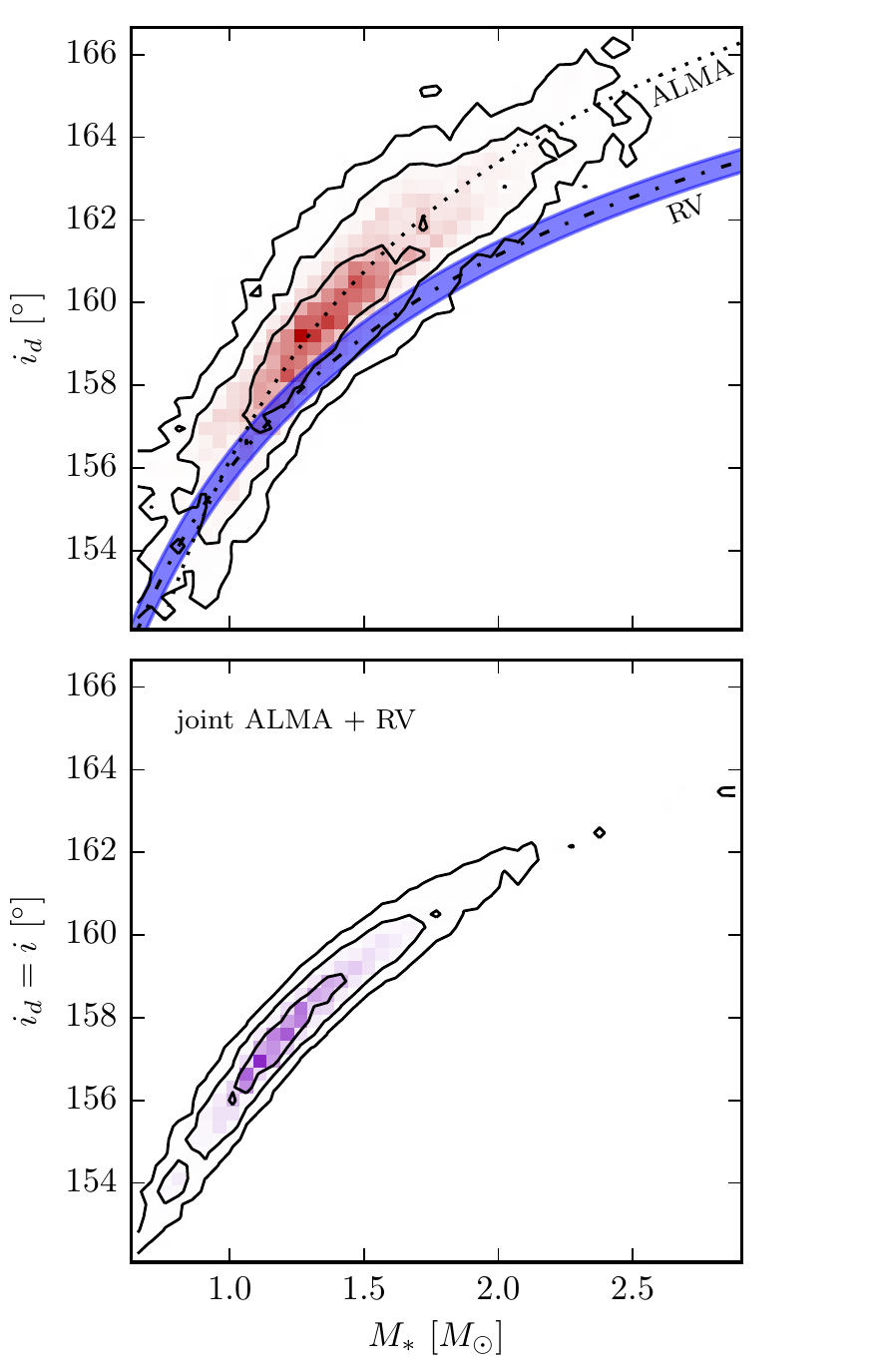}
  \figcaption{({\it top}): The joint posterior distribution for \{$M_\ast$, $i_d$\}, marginalized over all other parameters. To compare with the constraints from the updated binary orbit from RV monitoring measurements, we overlay ($\pm$1 $\sigma$) contours for the measurement of $M_\ast \sin^3 i$ (see Sect.~\ref{sec:orbit}).  ({\it bottom}): The joint posterior distribution combining the RV and disk measurements and assuming $i = i_d$. Contours denote 1, 2, and 3 $\sigma$ levels.
  \label{fig:triangle}}
\end{figure}

We explore the 12-dimensional posterior-space with an ensemble MCMC sampler \citep{foreman-mackey13}, employing uniform (uninformative) priors on all parameters except for $i_d$ and $d$. We adopt a standard geometrical prior on the disk inclination, $p(i_d)=\sin(i_d)/2$, reflecting that there are more disk orientations that result in edge-on than face-on viewing angles.\footnote{As we show below, the disk plane is near the plane of the sky.  Given that fact and the substantial cloud contamination near the systemic velocity, we cannot uniquely determine the direction of the angular momentum axis (i.e., the {\it sense} of $i_d$, whether it is $\sim$160\degr\ or $\sim$20\degr) from the ALMA data alone.  Therefore, we employ the astrometric constraint made by \citet{boden09} from infrared interferometry measurements to enforce $i > 90\degr$ in our analysis.}  We choose a conservative Gaussian prior on the distance with a mean of 145\,pc and a width ($\sigma$) of 20\,pc, meant to represent the range of possible distances to sources in the Taurus clouds \citep[e.g.,][]{torres10}.  In this analysis, we also conservatively exclude from the likelihood calculations 23 spectral channels that show evidence of molecular cloud contamination; these are marked in Fig.~\ref{fig:results}.  The resulting inferences on the model parameters are listed in Table~\ref{table:parameters}.  A comparison of the data with the model is shown in the form of channel maps in Figure~\ref{fig:results}.  As was demonstrated clearly in previous work \citep[e.g.,][]{simon00,rosenfeld12}, the density- and temperature-related parameters have negligible impact on an inference of the host mass.  The key parameters are $M_\ast$ and $i_d$: the \{$M_\ast$, $i_d$\} joint posterior distribution is shown in Figure~\ref{fig:triangle}.    

\capstartfalse
\begin{deluxetable}{lr@{ $\pm$ }l|lr@{ $\pm$ }l}
  \tablecaption{\label{table:parameters}Inferred Parameters for DQ Tau}
  \tablehead{\colhead{{\sc Parameter}} & \multicolumn{2}{c}{{\sc Value}} & \colhead{{\sc Parameter}} & \multicolumn{2}{c}{{\sc Value}}}
  \startdata
  $T_{10}$ (K)                  & 121    & 10   & $i_d$ (${}^\circ$)     &  160  & 3    \\
  $q$                           & 0.71   & 0.02 & $\varphi$ (${}^\circ$) & 94.2  & 0.5  \\
  $\log M_{\rm CO}$ ($M_\odot$) & $-$8.0 & 0.3  & $v_{\rm sys}$ ($\kms$) & $+$9.24  & 0.01 \\
  $r_c$ (AU)                    & 28     & 4    & $\Delta_\alpha$ (\arcsec) & $-$0.088 & 0.003 \\
  $M_\ast$ ($M_\odot$)          & \multicolumn{2}{c|}{$1.27_{-0.27}^{+0.46}$} & $\Delta_\delta$ (\arcsec) & $-$0.246 & 0.003 \\
  $\xi$ ($\kms$)                & 0.18   & 0.02 & $d$ (pc)               & 155   & 15 
  \enddata
  \tablecomments{The quoted uncertainties represent the maximum likelihood estimate and the 68.3\%\ highest density interval computed around this value. The systemic velocity is given in the LSRK frame for the standard radio definition, and corresponds to $+21.95 \pm 0.01\, \kms$ in the barycentric frame. Samples from the posterior are published at \url{https://figshare.com/articles/MCMC_Samples/2063424}. 
  }
\end{deluxetable}
\capstarttrue

We infer a mass of $1.27_{-0.27}^{+0.46}\,M_\odot$ for the DQ~Tau binary, marginalized over the uncertainty contained in our distance prior.  This can be expressed in a distance-independent manner as  $M_{\ast}/d = 0.0086_{-0.0018}^{+0.0021}\,M_{\odot}$ pc$^{-1}$; the formal uncertainty on $M_{\ast}$ is $\sim$25\%\ if the distance is known exactly.  This precision is significantly poorer than for most disk-based dynamical mass measurements, due to the unfortunate combination of a relatively face-on viewing geometry ($i_d \approx 160\degr$) and the cloud contamination around the systemic velocity.  

\subsection{An Updated Spectroscopic Orbital Solution} \label{sec:orbit}

The orbital solution by \cite{mathieu97} used their 14 pairs of CfA RVs along with the velocities measured from the Lick, Keck, and McDonald observatories (hereafter the `LKM' set). Because of difficulties in maintaining a consistent velocity zero point from night to night and instrument to instrument, the latter data were originally derived only as velocity \emph{differences} between the primary and secondary, rather than individual velocities for each star. To incorporate these LKM data into a conventional double-lined orbital solution, and at the same time to tie those observations to the CfA zero point, \cite{mathieu97} constructed primary and secondary ``measurements'' from each velocity difference. They did this assuming a fixed mass ratio (of unity) and center-of-mass velocity, based on the values inferred from an initial fit based only on their 14 CfA velocity pairs (where they found $M_2/M_1 = 0.97 \pm 0.15$ and $\gamma = 22$~$\kms$).  They then combined all of the measurements into a final fit, but were careful to note that both $\gamma$ and the velocity semiamplitudes $K_1$ and $K_2$ are biased and should be ignored in favor of the values from the CfA-only solution, and similarly with the minimum masses and projected semimajor axes, which depend on the semiamplitudes.

For this work we have preferred to incorporate the LKM velocity differences directly into the fit in their original form. We therefore reconstructed the original velocity differences trivially from the primary and secondary ``measurements'' reported by \cite{mathieu97} in their Table~1. The 22 RV differences were combined in a weighted least-squares fit with our 33 pairs of primary/secondary velocities, yielding the elements listed in Table~\ref{table:orbit}. For the individual velocities, weights were calculated from the internal errors. The LKM velocity differences were assigned reasonable nominal errors to begin with, and all uncertainties were then adjusted iteratively so as to obtain reduced $\chi^2$ values near unity for each type of measurement (primary, secondary, RV difference). Final root-mean-square residuals, which are representative of the typical measurement errors, are given in the table. The global fit derives most of the constraint on the mass ratio from the individual primary and secondary velocities. The RV differences strongly constrain the $K_1+K_2$ sum, but they also help to strengthen the individual $K$ values indirectly to some extent through constraints on the remaining orbital elements. We initially allowed for a difference in the center-of-mass velocities for the primary and secondary, to account for possible biases in the RVs that may occur as a result of template mismatch, but found the difference to be insignificantly different from zero ($-0.61 \pm 0.71$~$\kms$). Consequently, the final fit assumed a common value of $\gamma$.

\capstartfalse
\begin{deluxetable}{lc}
\tablewidth{0pc}
\tablecaption{Updated spectroscopic orbital solution for DQ~Tau.\label{table:orbit}}
\tablehead{
\colhead{\hfil~~~~~~~~~~~~~Parameter~~~~~~~~~~~~~} &
\colhead{Value}
}
\startdata
$P$ (days)\dotfill                         &    15.80158~$\pm$~0.00066\phn     \\
$\gamma$ ($\kms$)\dotfill                  &    $+$24.52~$\pm$~0.33\phn\phs    \\
$K_1$ ($\kms$)\dotfill                     &       20.28~$\pm$~0.71\phn        \\
$K_2$ ($\kms$)\dotfill                     &       21.66~$\pm$~0.60\phn        \\
$e$\dotfill                                &       0.568~$\pm$~0.013           \\
$\omega_1$ (deg)\dotfill                   &       231.9~$\pm$~1.8\phn\phn     \\
$T_{\rm peri}$ (HJD$-$2,400,000)\dotfill   &   47433.507~$\pm$~0.094\phm{2222} \\
$M_1 \sin^3 i$ ($M_{\sun}$)\dotfill        &      0.0348~$\pm$~0.0017          \\
$M_2 \sin^3 i$ ($M_{\sun}$)\dotfill        &      0.0326~$\pm$~0.0020          \\
$(M_1+M_2) \sin^3 i$ ($M_{\sun}$)\dotfill  &      0.0674~$\pm$~0.0033          \\
$a_1 \sin i$ ($10^6$ km)\dotfill           &        3.63~$\pm$~0.13            \\
$a_2 \sin i$ ($10^6$ km)\dotfill           &        3.87~$\pm$~0.11            \\
$a \sin i$ ($R_{\sun}$)\dotfill            &       10.78~$\pm$~0.18\phn        \\
$q \equiv M_2/M_1$\dotfill                 &       0.936~$\pm$~0.051           \\
$\sigma_1$ ($\kms$)\dotfill                &             3.44                  \\
$\sigma_2$ ($\kms$)\dotfill                &             2.26                  \\
$\sigma_{\rm LKM}$ ($\kms$)\dotfill        &             2.50                  \\
Time span (days)\dotfill                   &            10668.8                \\
Time span (orbital cycles)\dotfill         &             675.2                 \\
$N_{\rm RV}$\dotfill                       &         $33 \times 2$             \\
$N_{\rm LKM}$\dotfill                      &              22
\enddata
\tablecomments{These results are based on a joint fit of the individual primary/secondary velocities from CfA and the LKM velocity differences. The physical constants used here are those adopted by \cite{torres10}, consistent with the 2015 IAU Resolution B3.}
\end{deluxetable}
\capstarttrue

\begin{figure}
\epsscale{1.15}
\plotone{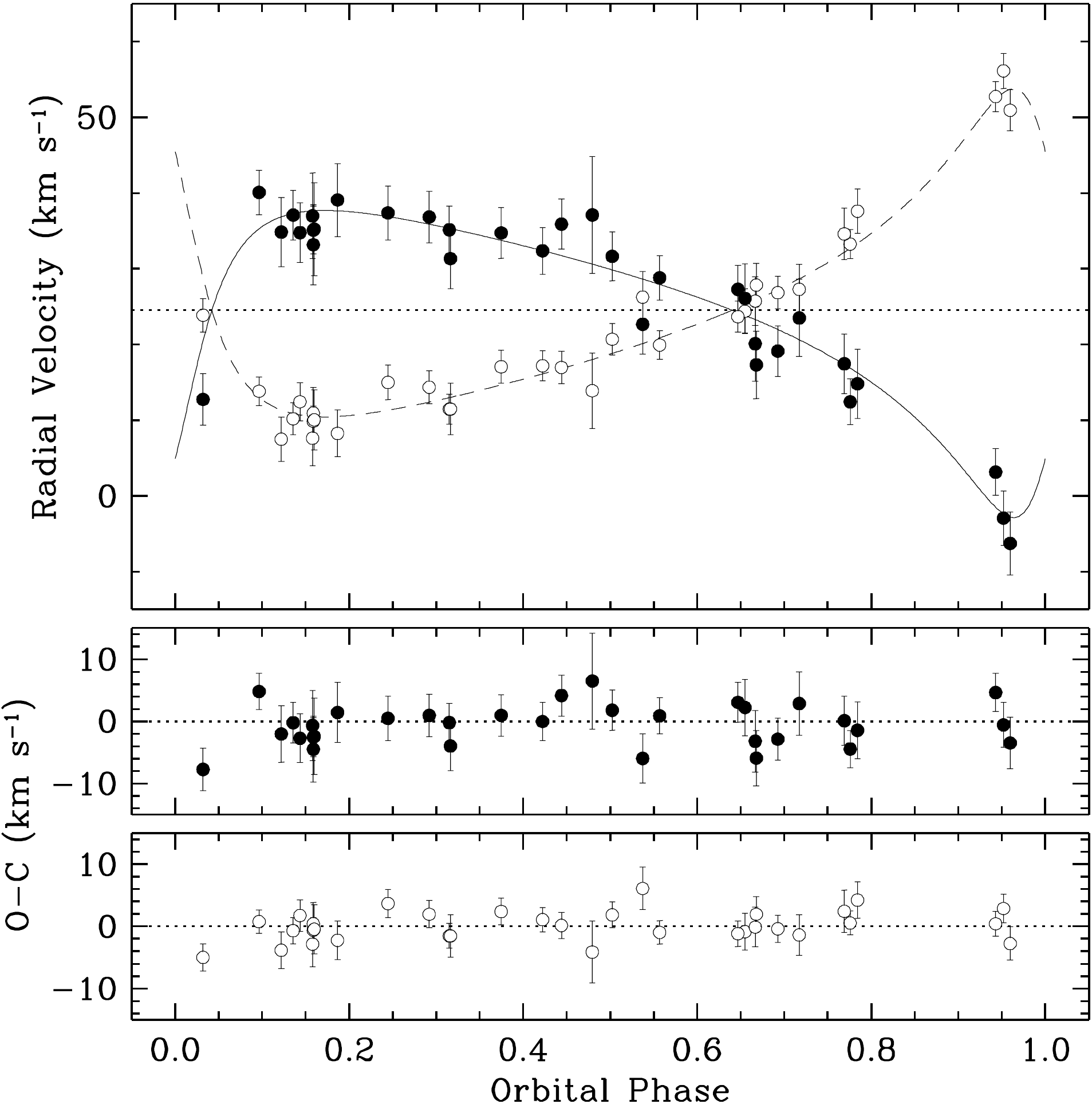}
\figcaption[]{CfA radial-velocity measurements of DQ~Tau as a function of orbital phase, including our best fit model that uses also the LKM velocity differences. Primary velocities are represented with filled symbols, and the dotted line marks the center-of-mass velocity. The bottom panels show the residuals. \label{fig:orbit}}
\end{figure}

\begin{figure}
\epsscale{1.15}
\plotone{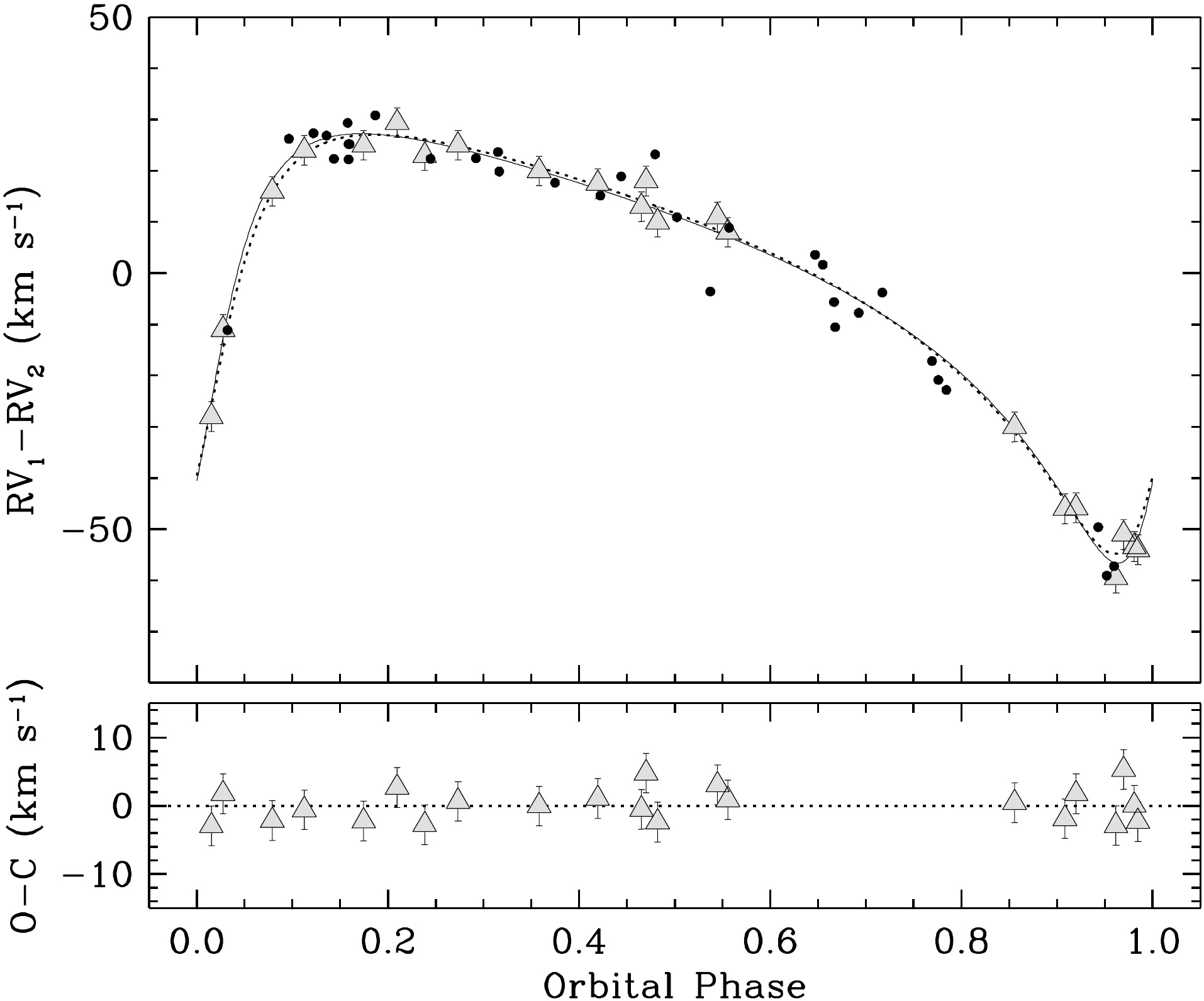}
\figcaption[]{Predicted velocity differences as a function of orbital phase according to our best-fit model (solid line), shown with the measured LKM values (triangles). Residuals are displayed at the bottom. Also shown for reference in the top panel is an orbit model that uses only the 33 individual RV measurements from CfA (dotted line), which is nearly indistinguishable from the global fit. The dots represent the velocity differences we compute from the CfA measurements, to show that both types of measurements are fully consistent with each other. \label{fig:RVdiff}}
\end{figure}

A graphical representation of the 33 pairs of primary/secondary velocities from CfA is presented in Figure~\ref{fig:orbit}, along with our best-fit model from the global fit that includes the LKM velocity differences. In Figure~\ref{fig:RVdiff} we illustrate the good agreement between the same best-fit model (solid curve) and the LKM velocity differences. The deviations between this best-fit model and a separate one that uses only the 33 individual CfA velocities (dotted line in the figure) are minimal.

Our results in Table~\ref{table:orbit} are generally consistent with those of \cite{mathieu97}, but with uncertainties typically reduced by factors of 2--5. The minimum masses now have relative uncertainties of 5--6\% instead of $\sim$21\%.

\section{Discussion}\label{sec:disc}

We have reported a new dynamical constraint on the mass of the young DQ~Tau binary made by reconstructing the velocity field of its circumbinary disk using ALMA observations of its CO line emission, as well as an update on the binary orbital parameters based on a long-term optical spectroscopic monitoring campaign.  In the following sections, we compare these constraints in more detail and discuss them in the context of pre-MS evolutionary model predictions and similar measurements for other equal-mass binary systems.

\subsection{Comparison of Disk- and Binary-based Constraints} \label{subsec:rv}

The disk-based dynamical mass approach formally constrains the quantity $(M_\ast / d)\, \sin^2 i_d$ by reconstructing the sky-projected Keplerian velocity field of the gas disk.  Given some prior information on the distance and sufficient resolution to determine the aspect ratio of the emission ($i_d$), a precise estimate of $M_\ast$ can be made.  This is not quite the case for DQ~Tau.  The disk orbital plane is oriented such that it is viewed nearly in the plane of the sky, which concentrates most of the more spatially extended molecular line emission near the systemic velocity.  That would be fine given our ALMA observations, except for the ambient molecular cloud material that also produces extended emission at those same velocities.  Taken together, the small projection factor and severe cloud contamination significantly expand the \{$M_\ast$, $\sin^2 i_d$\} degeneracy (see Fig.~\ref{fig:triangle}), limiting the precision of our dynamical mass constraint.  For a conservative prior on $d$, we measure a joint constraint of $M_\ast\, \sin^2 i_d = 0.164\pm0.016\,M_\odot$, or individual measurements of $1.27_{-0.27}^{+0.46}\,M_\odot$ and $i_d = 160\pm3\degr$.  

The orbital solution for a double-lined spectroscopic binary determines $M_\ast\, \sin^3 i$ (independent of $d$) from a fit to a time series of RV measurements.  The updated solution presented here has $M_\ast \sin^3 i = 0.0674~\pm~0.0033\,M_\odot$. Figure~\ref{fig:triangle} confirms that the ALMA disk-based and RV binary-based constraints are in good agreement (well within 1\,$\sigma$) in the binary mass--inclination plane.  This suggests that these constraints can be combined together to yield some informative combined measurements for the system.  If we assume that the binary and disk orbital planes are exactly aligned ($i = i_d$), the joint constraints from the RV and ALMA data indicate $M_\ast = 1.21\pm0.26\,M_\odot$ and $i = 158\pm2\degr$ (this composite posterior is shown in the bottom panel of Fig.~\ref{fig:triangle}).  If we consider the ALMA constraint on the quantity $M_\ast/d$ rather than marginalizing over the prior on $d$, we can use the assumption of coplanarity and the RV data to estimate a dynamical distance to the system.  In that case, we estimate $d_{\rm dyn}= 184\pm26$\,pc, which, although imprecise, has a most probable value slightly higher (at the $\sim$1\,$\sigma$ level) than standard measurements for the Taurus star-forming region \citep[e.g.,][]{torres10} and our adopted prior, which may suggest a larger depth of the Taurus complex.  In the context of our nominal prior on $d$, we can also use both datasets to instead infer a limit on the mutual inclination angle between the disk and binary orbital planes: we find that $\psi \equiv i - i_d = -1.3 \pm 1.1\degr$. Interestingly, we note a small discrepancy between the systemic velocity in the barycentric frame derived from the disk ($+21.95 \pm 0.01\, \kms$) and that derived from the binary orbit ($+$24.52~$\pm$~0.33 $\kms$). We speculate that this offset may be caused by veiling of the stellar photospheres, which results in a sub-optimal fit of the spectroscopic templates used for the radial velocity determinations.

\subsection{Comparison to Pre-MS Evolution Models} \label{subsec:evol_models}
Having demonstrated that independent dynamical constraints on $M_\ast$ for the DQ~Tau binary yield consistent results, it is of interest to make a comparison with the more common approach of estimating masses (and ages) from theoretical pre-MS evolutionary models.  

A range of (combined-light) spectral types have been reported for DQ~Tau, with a general consensus around M0--M1.  Individual spectral diagnostics often skew towards earlier or later spectral types: for example, \citet{basri97} found that ratios of temperature sensitive \ion{Sc}{1} lines suggest a K4--K5 classification, while \citet{bary14} showed that many infrared molecular features (e.g., TiO, FeH, and H$_2$O) are better matched with a later type, M2.5--M4.5.  Some of this ambiguity may be due to the implicit assumption that both stars have identical photospheric properties.  The improved orbital solution in Sect.~\ref{sec:orbit} suggests otherwise: the inferred mass ratio ($M_2/M_1 = 0.93\pm0.05$) indicates that the DQ~Tau stars have different temperatures and luminosities.    

With that in mind, we explored a two-component fit to the $BVRIJ$ photometry compiled by \citet{rydgren84}, \citet{kolotilov89}, and \citet{skrutskie06} \citep[previously presented by][]{andrews13}.  Observations in the $U$-band and at longer infrared wavelengths were excluded due to contamination by accretion activity and dust emission, respectively.  The adopted model magnitudes were interpolated for a given \{$T_{\rm eff}$, $\log g$\} from the {\sc BT-Settl} synthetic photometry catalog \citep{allard03} for solar metallicity.  These were adjusted for extinction using the \citet{fitzpatrick99} reddening law (with $R_V = 3.1$) and scaled to account for a given luminosity (assuming the same prior on $d$ as in Sect.~\ref{sec:disk}).  After some experimentation, we found that the effects of surface gravity are relatively small (given the other uncertainties), so we fixed $\log g = 4.0$ for both stars.  Each model therefore has five physical parameters, \{$T_1$, $L_1$, $T_2$, $L_2$, $A_V$\}.  We used an additional five nuisance parameters (one per band) to describe the ``jitter" (dispersion) in each photometric band due to variability (presumed to be described by a Gaussian with mean zero and this parametric description of the variance).  The model quality for a given set of parameters was determined with a $\chi^2$ likelihood function and a reasonable set of priors.  At each posterior draw, we calculated the implied mass ratio and imposed a Gaussian prior with mean 0.93 and dispersion 0.05, based on the RV orbital solution.  Since photometry alone is a poor diagnostic of $T_{\rm eff}$ (especially for a composite dataset), we adopted a Gaussian prior with mean of 3900\,K and dispersion of 250\,K on $T_1$ and $T_2$ and enforced the conditions $T_2 \le T_1$ and $L_2 \le L_1$. 

\begin{figure}[ht!]
  \includegraphics{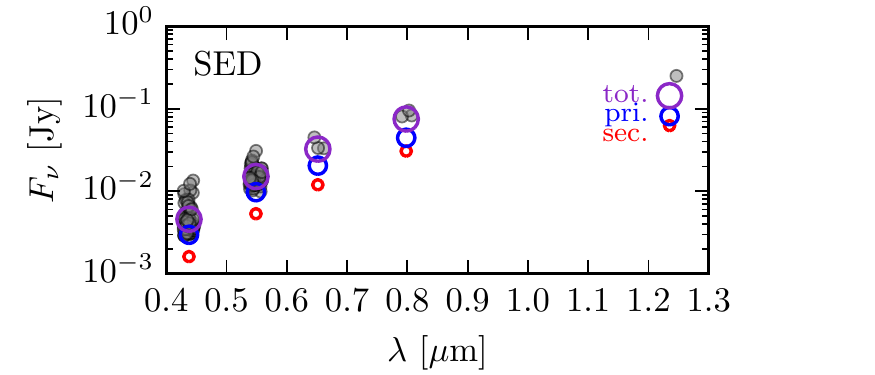}
  \includegraphics{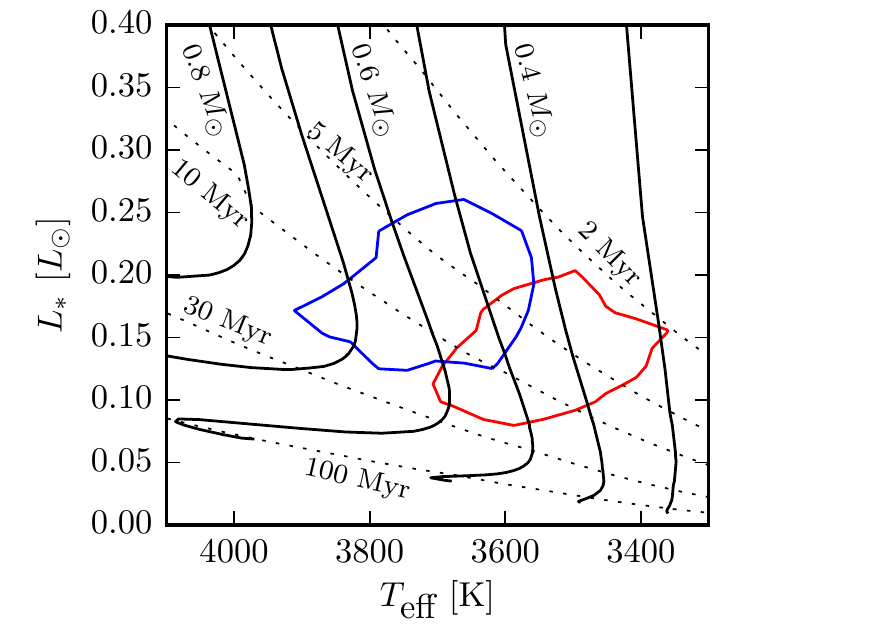}
  \includegraphics{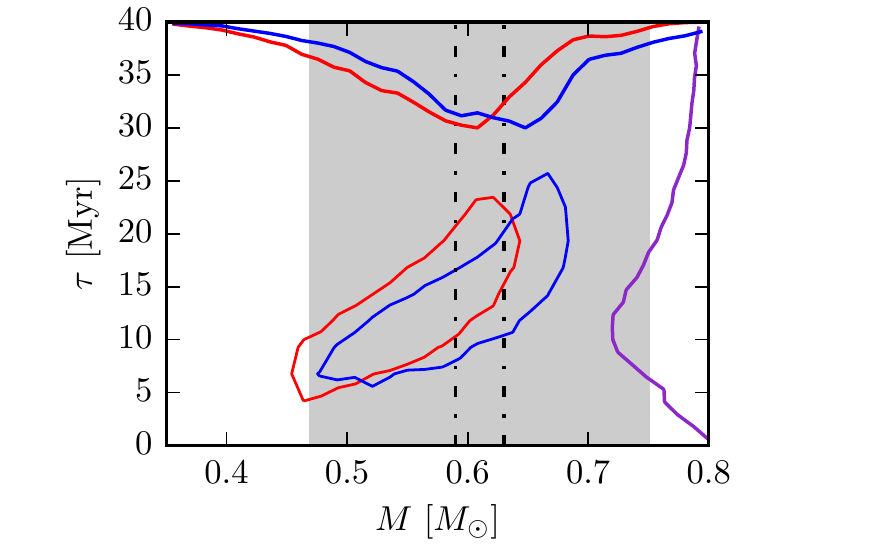}
  \figcaption{
  ({\it top}): The best-fit models of the broadband photometry overlaid on the data.  ({\it middle}) The resulting HR diagram, with the marginal posteriors inferred from the photometry modeling shown as 1\,$\sigma$ contours.  The \citet{dotter08} pre-MS model mass tracks and isochrones are overlaid.  ({\it bottom}): The joint mass and age constraints from the \citet{dotter08} pre-MS models assuming the stars are coeval, shown as 1\,$\sigma$ contours. The marginalized distributions are shown at the boundaries of the plot.  The gray band marks the disk-based constraint on the individual component mass $M_{\ast}$ (1\,$\sigma$). Other pre-MS model predictions give generally comparable results.
  \label{fig:PMS}}
\end{figure}

Figure~\ref{fig:PMS} provides a summary of the modeling results.  We find that the photometry prefers relatively low temperatures, $T_1 = 3700\pm200$\,K and $T_2 = 3500\pm175$\,K, and extinction, $A_V = 0.5 \pm 0.2$, yielding logarithmic luminosities $\log_{10} L/L_\odot$ of $-0.73 \pm 0.16 $ and $-0.87 \pm 0.16$ for the primary and secondary, respectively.  These are not particularly stringent constraints on the binary location in the HR diagram, of course, owing to the relatively ambiguous spectral classifications available for the individual components.

Based on the joint posterior distribution of \{$T_1$, $L_1$, $T_2$, $L_2$\}, we followed the formalism of \citet{jorgensen05} to derive component masses and ages \{$\tau$, $M_1$, $M_2$\} from the predictions of pre-MS evolutionary models in the HR diagram, assuming that the binary stars are coeval.  Various incarnations of such models \citep{siess00,dotter08,tognelli11,baraffe15} make consistent predictions within the (considerable) uncertainties, indicating a total binary mass of $1.20\pm0.16\,M_\odot$ that is in good agreement with the dynamical constraints from the ALMA and RV data.  The corresponding age predictions are considerably more uncertain; favored values are in the 6--10\,Myr range, although the permissible ages span from $\sim$6 to 20\,Myr (1\,$\sigma$). We note that this analysis is under the assumption of coevality of the two stars, which may not necessarily be true. Additionally, the unusual nature of the DQ~Tau system (e.g., colliding magnetospheres during periastron) may also invalidate our assumptions of normal pre-main sequence evolution.

\section{Summary and Context} \label{sec:summary}

We have presented a set of new constraints on the fundamental properties of the DQ~Tau young binary system, based on ALMA observations of molecular line emission from its circumbinary disk and an updated analysis of optical spectroscopic measurements of its (stellar) radial velocity variations.  For a conservative distance prior ($d = 145\pm20$\,pc), we find that the disk-based and binary-based {\it dynamical} constraints on the total stellar mass in the DQ~Tau system are in excellent agreement: their combined inputs suggest a total mass $M_{\ast} = 1.21\pm0.26\,M_\odot$, and therefore individual component masses $M_1 = 0.63\pm0.13\,M_\odot$ and $M_2 = 0.59\pm0.13\,M_\odot$ (incorporating the uncertainty on $q$). Moreover, we also find that the disk and binary orbital planes are aligned within 3\degr, showing that the system is coplanar across radial distances from $\sim0.1$\,AU to 100\,AU. In this system, the dynamical mass precision is limited by an unfortunate combination of two factors: an orbital plane that is oriented nearly in the sky plane, and some large-scale contamination of the disk CO spectral emission from the ambient molecular cloud. In the future, an accurate parallax from \emph{GAIA} will help improve the precision of the disk-based estimate of $M_\ast$.

We also estimated the stellar mass in the system using the more common technique that compares the component locations in the HR diagram with predictions of theoretical pre-MS evolution models, and generally found good agreement.  However, that approach has restricted utility given the lack of component-resolved photometry and substantial ambiguity on the effective temperatures.  There is still much to be learned from this fascinating system; our mass constraints lend some quantitative benchmarks that can be adopted in future studies.  

DQ~Tau is the third nearly-equal mass young binary system that has been analyzed with these two {\it independent} dynamical techniques to constrain stellar masses, the others being the older and earlier type systems V4046~Sgr \citep[total mass $1.75\,M_\odot$;][]{rosenfeld12} and AK~Sco \citep[total mass $2.50\,M_\odot$;][]{czekala15}.  Using millimeter-wave interferometric measurements of their CO spectral line emission, model fitting that reconstructs the Keplerian velocity fields of their circumbinary disks finds dynamical masses that are in excellent agreement with constraints from optical RV monitoring of the host binaries (thereby also implying that the binary and disk orbits are co-planar).  Granted, this is a small sample, but it does span an important range of system properties: e.g., spectral types from early M to mid F, ages from a few to tens of Myr, and orbital eccentricities from circular to $e \approx 0.6$.  Taken together, this work validates the quantitative {\it accuracy} of the disk-based dynamical inference of young star masses, provided that it is done carefully in a proper analysis framework. Moving forward, this confirms that ALMA should play a substantial role in young star astrophysics, as the technique used here is the only means of precisely measuring the masses for large samples of {\it single} stars.

\acknowledgments
We appreciate some useful computational suggestions from Ryan Loomis.  IC gratefully acknowledges funding support from the Smithsonian Institution.  SA appreciates the very helpful support provided by the NRAO Student Observing Support program related to the early development of this project.  This paper makes use of the following ALMA data: 2012.1.00496.S.  ALMA is a partnership of ESO (representing its member states), NSF (USA), and NINS (Japan), together with NRC (Canada) and NSC and ASIAA (Taiwan), in cooperation with the Republic of Chile.  The Joint ALMA Observatory is operated by ESO, AUI/NRAO, and NAOJ.  Figure~\ref{fig:triangle} was generated with the \texttt{triangle.py} code \citep{foreman-mackey14}. This research made extensive use of the Julia programming language \citep{julia12} and Astropy software package \citep{astropy13}.

\end{document}